\documentclass[
reprint,
superscriptaddress,
showpacs,
aps,
prb,
]{revtex4-1}

\usepackage{amsmath}
\usepackage{amssymb}%
\usepackage{natbib}
\usepackage{graphicx}
\usepackage{dcolumn}
\usepackage{bm}
\usepackage{ulem,color}

\newcommand{\lag}{\langle}
\newcommand{\rag}{\rangle}

\newcommand{\hspc}{\hspace{1em}}
\newcommand{\htab}{\hspace{2em}}

\newcommand{\mrm}{\mathrm}
\newcommand{\mcl}{\mathcal}

\begin{document}

\title{Magnetization Plateaux by Reconstructed Quasi-spinons \\in a Frustrated Two-Leg Spin Ladder under a Magnetic Field}

\author{Takanori Sugimoto}
\email{sugimoto.takanori@rs.tus.ac.jp}
\affiliation{Department of Applied Physics, Tokyo University of Science, Katsushika, Tokyo 125-8585, Japan}
\author{Michiyasu Mori}
\affiliation{Advanced Science Research Center, Japan Atomic Energy Agency, Tokai, Ibaraki, 319-1195, Japan}
\author{Takami Tohyama}
\affiliation{Department of Applied Physics, Tokyo University of Science, Katsushika, Tokyo 125-8585, Japan}
\author{Sadamichi Maekawa}
\affiliation{Advanced Science Research Center, Japan Atomic Energy Agency, Tokai, Ibaraki, 319-1195, Japan}

\date{\today}

\begin{abstract}
The quantum phase transitions induced by a magnetic field are theoretically studied in a frustrated two-leg spin ladder.
Using the density-matrix renormalization-group method, we find some magnetic phase transitions and plateaux in two different cases of strong and weak rung couplings.
With the strong rung coupling, the three magnetization plateaux are found at 1/3, 1/2, and 2/3 due to the frustration. 
Those can be understood in terms of a quasi-spinon reconstructed from the singlet and the triplets of spins on a rung. 
The plateau at 1/2 corresponds to the valence bond solid of the quasi-spinons, while the plateaux at 1/3 and 2/3 can be associated with the array of quasi-spinons such as soliton lattice. 
This is different from the usual Bose-Einstein-condensation picture of triplons. 
Our results will be useful to understand magnetization curves in BiCu$_2$PO$_6$.
\end{abstract}


\maketitle

\section{Introduction}
\label{sec:intro}
Correlated many-body systems have rich variety of quantum phenomena such as Mott insulator, high temperature superconductivity, Kondo effect, fractional quantum Hall effect and so on~\cite{maekawa04,fulde12}. 
Interacting boson systems also have many interesting physics such as superfluidity accompanied by the Bose-Einstein condensation (BEC). 
A quantum spin system is useful for studying such an interacting boson system due to a mapping between interacting spins- and bosons-systems~\cite{matsubara56}. 
In fact, the BEC in quantum anti-ferromagnet TlCuCl$_3$ has been studied experimentally~\cite{nikuni00,ruegg03} and theoretically~\cite{giamarchi99,giamarchi08}. 
Moreover, many quantum magnets have been examined so far~\cite{zapf14}. 

The density of bosons in quantum magnets is tuned by an applied magnetic field, which induces the BEC corresponding to the long-ranged magnetic order. It can be seen in a magnetization curve. 
For example, in the dimerized quantum anti-ferromagnet TlCuCl$_3$, which has a gap in the spin excitation, the magnetic field has a critical value, $H_{c1}$, where the gap is destroyed and the magnetization becomes finite. 
Such a behavior is explained by the BEC of triplons~\cite{nikuni00,giamarchi99,zapf14}, which are triplet states on a dimer and are associated with hard core bosons. 
Near $H\gtrsim H_{c1}$, the density of triplons is still dilute and will make some kinds of bound states~\cite{sushkov98,kotov99}. Increasing the magnetic field, $H$, the magnetization will become constant above a saturation field, $H_{c2}$.
In $H_{c1}<H<H_{c2}$, strong repulsion between triplons on a lattice can induce magnetization plateaux, which corresponds to Mott insulator phase of bosons~\cite{hida94,oshikawa97,totsuka98,okunishi03}.

The triplon BEC mentioned above is accepted as a good starting point of understanding magnetization plateaux. In this paper, we proposed a new concept for magnetization plateaux, which is different from the triplon picture. Such a new concept works on a frustrated two-leg spin ladder (F-2LSL). Calculating magnetization curves of the F-2LSL by the density-matrix renormalization-group (DMRG) method, we find three fractional magnetization plateaux due to frustration in the case of strong rung coupling. We interpret the origin of the plateaux by introducing a quasi-spinon constructed by a singlet and a triplet of a spin pair on a rung. 
In contrast to the triplon BEC picture, the magnetization plateaux correspond to the valence bond solid and the solitonic lattice of the quasi-spinons.
As a realistic material containing the F-2LSL, BiCu$_2$PO$_6$ (BCPO) has been actively studied~\cite{tsirlin10,kohama12,lavarelo12,casola13,sugimoto13,kohama14,sugimoto14}. 
Casola {\it et al}. have reported that a field-induced phase can be fit by a solitonic excitation of spinons originating from triplons instead of the triplon BEC picture~\cite{casola13}, indicating a reconstruction of the quasi-particles and field-induced second-order phase transition. Our new concept of quasi-spinons will give a hint for fully understanding the field-induced phases of BCPO as well as other frustrated spin chains that exhibit fractional magnetization plateaux~\cite{hida94,oshikawa97,totsuka98,okunishi03}.

The rest of the paper is organized as follows. 
In Sec.~\ref{sec:model}, the Hamiltonian of the frustrated two-leg spin ladder will be defined, and a way to calculate the magnetization curves using the DMRG method will be explained.
In Sec.~\ref{sec:res}, numerical results of the magnetization curves, which  depend on both the rung couplings and the frustration, will be shown in detail. 
We will formulate quasi-spin operators to explicitly derive an  effective Hamiltonian with the strong frustration and the strong rung couplings in an applied magnetic field. 
How to interpret the magnetization plateaux in terms of the effective Hamiltonian is also discussed together with schematic pictures. Relation between weak and strong rung-coupling limits will be summarized in a table.
Finally, summary and discussions will be given in Sec.~\ref{sec:sum}.

\section{Model and Method}\label{sec:model}
A model Hamiltonian of the F-2LSL with an applied magnetic field $H$ is given by
\begin{equation}
\mcl{H}=\mcl{H}_\parallel+\mcl{H}_\perp+\mcl{H}_{Z}
\label{eq:ham0}
\end{equation}
with
\begin{eqnarray}
\mcl{H}_\parallel&=&\sum_{\eta=1,2} J_{\eta} \sum_j\sum_{i=\mrm{u,l}}\bm{S}_{j,i}\cdot\bm{S}_{j+\eta,i},\\
\mcl{H}_\perp&=&J_\perp\sum_j\bm{S}_{j,\mrm{u}}\cdot\bm{S}_{j,\mrm{l}}, \\
\mcl{H}_{Z}&=& - H \sum_{j} \sum_{i=\mrm{u,l}} {S}_{j,i}^z,
\end{eqnarray}
where $\bm{S}_{j,\mrm{u(l)}}$ is the $S=1/2$ spin operator on the $j$ site in the upper (lower) chain and its $z$ component is $S^z_{j,\mrm{u(l)}}$.
There are three types of anti-ferromagnetic Heisenberg interactions, a nearest-neighbor coupling $J_1$ and a next-nearest-neighbor coupling $J_2$ in the leg direction, and a nearest-neighbor coupling on a rung bond $J_\perp$.
In the limit of weak rung coupling, $J_\perp\ll J_1 $, this model describes decoupled two frustrated spin chains, 
while a non-frustrated spin ladder is obtained in the limit of weak frustration, $J_2 \ll J_1$.
Consequently, this model bridges between a frustrated spin chain and a non-frustrated spin ladder through $J_\perp$ and $J_2$.

The preceding studies on this model have shown that there are two different phases as the ground state, i.e., the columnar-dimer and rung-singlet phases~\cite{lavarelo12}. In the columnar-dimer phase, the ground state is composed of two degenerated states with spontaneously-broken translational symmetry~\cite{majumdar69}, while no degeneracy exists in the ground state of the rung-singlet phase. In the former case, spinon as a fermionic quasi-particle helps us to understand the magnetic behavior, although the hard-core bosonic particle, triplon, gives a better explanation in the latter one. 
We note that this is one example of reconstruction of quasi-particles caused by the second-order phase transition with spontaneously-broken symmetry.

We calculate the magnetization curves in both the weak- and strong-rung-coupling limits by using the DMRG method with an open boundary condition~\cite{note1}.
First, we calculate the minimum energies $E_m$ with fixed magnetizations  $m=0,1,\dots,M_{\mrm{sat}}$ without a magnetic field. 
Then, the energy difference between $E_m$ and $E_{m+1}$ determines a magnitude of a magnetic field, 
\begin{equation}
	H_{m:m+1}=E_{m+1}-E_{m}, 
\end{equation}
at which the magnetization of the ground state changes from $m$ to $m+1$.  
Therefore, using $H_{m:m+1}$ and magnetic field $H$, the magnetiztion curve $M(H)$  is caluclated by, 
\begin{equation}
M(H)=\sum_{m=1}^{M_{\mrm{sat}}-1} m \, \theta(H-H_{m-1:m})\,\theta(H_{m:m+1}-H),
\end{equation}
where the $\theta(x)$ is the Heaviside step function.

\section{Numerical results and quasi-spin transformation}\label{sec:res}
\begin{figure*}
	\centering
	\includegraphics[width=\textwidth]{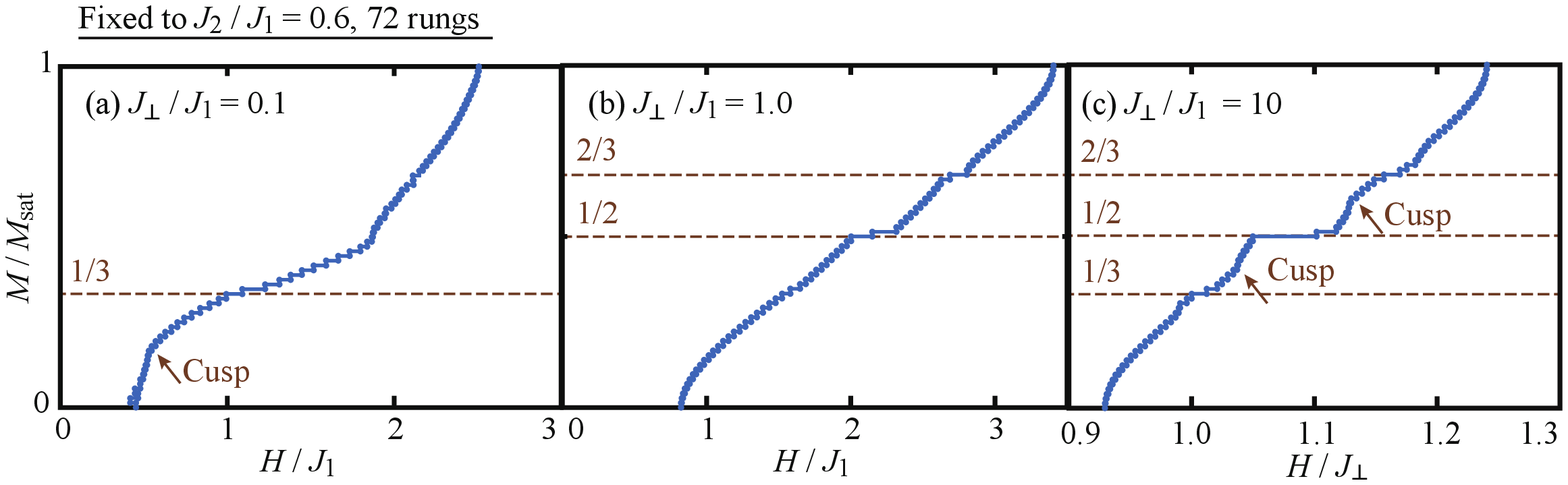}
	\includegraphics[width=\textwidth]{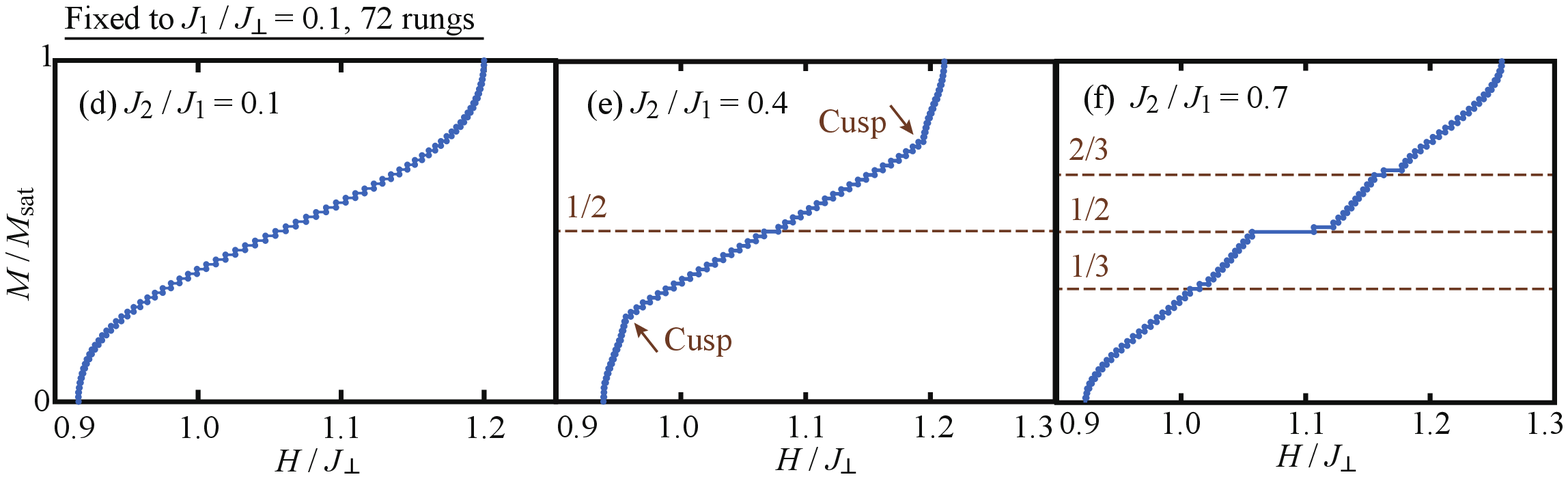}
	\caption{(Color online) Magnetization curves in a 72-rung F-2LSL. (a)-(c) Rung-coupling ($J_\perp/J_1$) dependence with a fixed frustration $J_2/J_1=0.6$.  (d)-(f) Frustration ($J_2/J_1$) dependence with a fixed rung coupling $J_1/J_\perp=0.1$. The one-third plateau denoted by ``1/3'' and the cusp singularity denoted by ``Cusp'' in (a) correspond to those of a frustrated spin chain. Other plateaux denoted by ``1/2'' and ``1/3'' appear at $J_\perp/J_1=1.0$ (b) and $J_1/J_\perp=0.1$ (c).
		The 1/3 plateau reappears in (c) together with some cusp singularities.
		In (d) there are no plateau and cusp, but in (e) two cusp singularities and 1/2 plateau appear. Strong frustration at $J_2/J_1=0.7$ causes three plateaux in (f).}
	\label{fig:mh}
\end{figure*}
\begin{figure*}
\centering
\includegraphics[width=0.8\textwidth]{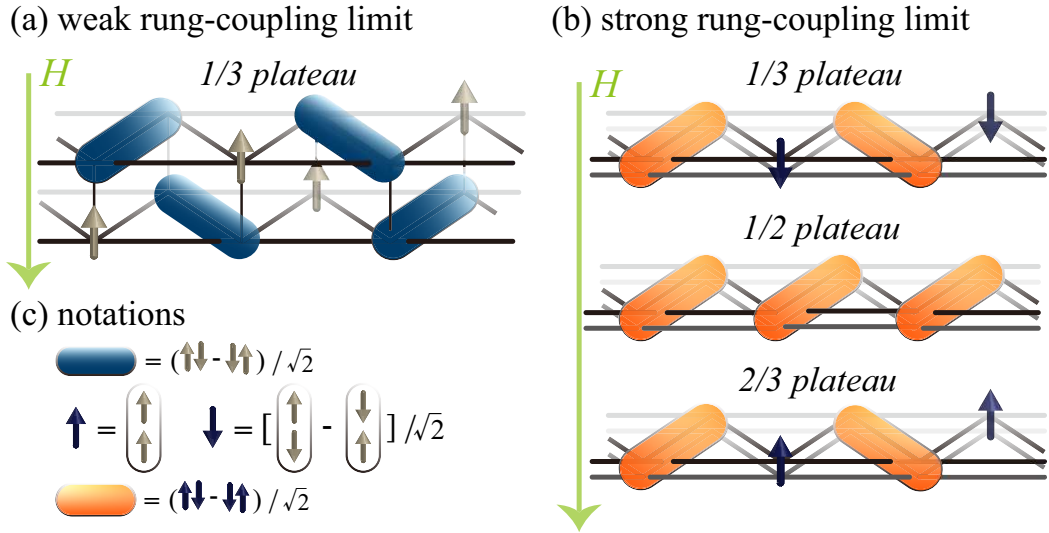}
\caption{(Color online) The schematic spin configurations for the plateau phases.
(a) Weak rung-coupling limit and (b) strong rung-coupling limit. (c) Schematic configurations used in (a) and (b). In (a), up spins locally freeze with long-ranged order and total $S^z$ in each three neighboring spins in the upper (lower) chain equals to $1/2$.
In (b), similar quasi-up-spin freezing with long-ranged order occurs in the $1/3$ and $2/3$ plateau phases, although quasi-spins play a physical role instead of the real spins.}
\label{fig:plateaux}
\end{figure*}

Figures \ref{fig:mh}(a)-(c) show that magnetization curves for rung couplings $J_\perp/J_1=0.1$, 1, and 10, with a fixed frustration $J_2/J_1=0.6$ in a 72-rung system.
In the weak rung-coupling limit, $J_\perp/J_1=0.1$, as shown in Fig.~\ref{fig:mh}(a), we can find a cusp singularity and a plateau at magnetization ratio $M/M_{\mrm{sat}}=1/3$, which are also reported in a frustrated chain~\cite{okunishi03}.
With an applied magnetic field, a three-fold degeneracy of triplet excitation breaks and one state of them goes down to the singlet ground state.
When the lowest energy of the triplets reaches down to that of the singlet ground state at $H=H_{c1}$, the gap-to-gapless transition occurs and the magnetization becomes finite.
A cusp singularity appears as a result of Lifshitz transition of Jordan-Wigner fermion~\cite{sugimoto14}.
The strong frustration induces a one-third plateau originated from a reconstructed unit cell with the size three times larger than that of an original unit cell.
Freezing spins with a magnetic moment make a long-ranged order in the leg direction together with deconfinement of spins between two legs (see Fig.~\ref{fig:plateaux}(a)).
Increasing the rung coupling up to $J_\perp/J_1=1$ in Fig.~\ref{fig:mh}(b), the one-third plateau is weakened or disappears, although other plateaux emerge at one-half magnetization $M/M_{\mrm{sat}}=1/2$ and two-third one $M/M_{\mrm{sat}}=2/3$.
Moreover, with a strong rung coupling in Fig.~\ref{fig:mh}(c), we can find again the one-third plateau with some cusp singularities.
The three plateaux at magnetization ratios $M/M_{\mrm{sat}}=1/3$, 1/2 and 2/3 does not satisfy the condition proposed by Oshikawa {\it et.~al.}~\cite{oshikawa97}, which is given by,  $Q(S-m^z)\in\mathbb{Z}$, with spin $S$, periodicity of lattice $Q$ ($Q=2$ for our Hamiltonian), magnetization $m^z=MS/M_{\mrm{sat}}$ at plateaux, and the set of integers $\mathbb{Z}$. 
In accordance with the Oshikawa-Yamanaka-Affleck criteria, these are caused by spontaneously-broken symmetries of the plateau states in the present system.
Since such a plateau state requires a confirmation with other methods like the numerical calculation and a bosonization approach one by one, these could be non-trivial, although potential plateaux and a possible case of spontaneously-broken symmetries have been discussed in several quantum spin models with similar characteristics~\cite{oshikawa97,totsuka98,cabra00,note2}.

The frustration dependence of the magnetization curve with the strong rung coupling $J_\perp/J_1=10$ is also calculated with $J_2/J_1= 0.1$, 0.4, and 0.7 in Figs.~\ref{fig:mh}(d), (e), and (f), respectively.
It is clear that cusp singularities and plateaux are caused by the frustration controlled by $J_2/J_1$, even though the number of singularities and plateau is different from the weak rung-coupling cases shown in Figs.~\ref{fig:mh}(a)-(c).
The spin freezing and long-ranged order with the one-third plateau state appear at $J_2/J_1= 0.7$.
In contrast to one-third plateau for weak-rung coupling, the periodicity of the freezing spins must be identical between the upper and lower chains because of the strong-rung coupling (see Fig.~\ref{fig:plateaux}(b)).

In order to clarify the origin of the magnetization plateaux, a quasi-spin transformation~\cite{giamarchi99} is applied into the original Hamiltonian (\ref{eq:ham0}) with a strong rung coupling.
In this limit, the hard-core-bosonic picture is still applicable up to $H\sim H_{c1}$.
However, the picture becomes worse in such a large magnetic field, because the symmetry of the triplets is no longer alive.
Instead of the symmetry of the triplets, a new SU(2) symmetry constructed by a singlet and a triplet appears with a quasi-spin transformation given by,
\begin{eqnarray}
T_{j,\mrm{p}}^\pm&=&\pm\frac{i}{\sqrt{2}}\left[S_{j,\mrm{u}}^\pm e^{\pm i\frac{\pi}{2}\left(S_{j,\mrm{l}}^z+\frac{1}{2}\right)}+S_{j,\mrm{l}}^\pm e^{\mp i\frac{\pi}{2}\left(S_{j,\mrm{u}}^z+\frac{1}{2}\right)}\right], \nonumber\\
T_{j,\mrm{p}}^z&=&\frac{1}{2}\left[(S_{j,\mrm{u}}^z+S_{j,\mrm{l}}^z)+S_{j,\mrm{u}}^+S_{j,\mrm{l}}^-+S_{j,\mrm{u}}^-S_{j,\mrm{l}}^+\right], \label{eq:top1}
\end{eqnarray}
and
\begin{eqnarray}
T_{j,\mrm{m}}^\pm&=&\frac{1}{\sqrt{2}}\left[S_{j,\mrm{u}}^\mp e^{\pm i\frac{\pi}{2}\left(S_{j,\mrm{l}}^z+\frac{1}{2}\right)}-S_{j,\mrm{l}}^\mp e^{\mp i\frac{\pi}{2}\left(S_{j,\mrm{u}}^z+\frac{1}{2}\right)}\right],\nonumber\\
T_{j,\mrm{m}}^z&=&\frac{1}{2}\left[-(S_{j,\mrm{u}}^z+S_{j,\mrm{l}}^z)+S_{j,\mrm{u}}^+S_{j,\mrm{l}}^-+S_{j,\mrm{u}}^-S_{j,\mrm{l}}^+\right]. \label{eq:top2}
\end{eqnarray}

With the transformation, $T$ operators obey the spin SU(2) algebra: $\{T_{j,\alpha}^+,T_{j,\alpha}^-\}=1$ and $[T_{i,\alpha}^+,T_{j,\alpha}^-]=0$ for $i\neq j$ and $\alpha=\mrm{p},\mrm{m}$.
Therefore, the transformation makes us possible to consider another SU(2) quasi-spin model of two different energy scale discussed below.
The original Hamiltonian with no perturbations ($J_1=J_2=0$) is rewritten as,
\begin{equation}
\mcl{H}_\perp + \mcl{H}_Z = \sum_j \mu_{j,\mrm{p}} n_{j,\mrm{p}} + \sum_j \mu_{j,\mrm{m}} n_{j,\mrm{m}} + \mrm{const.},
\end{equation}
where $n_{j,\alpha}(=T_{j,\alpha}^z+1/2)$ is number operators of the quasi-spins. 
The effective chemical potentials for the quasi-spins are given by $\mu_{j,\mrm{p}} =J_\perp-H $ and $\mu_{j,\mrm{m}}=J_\perp\left(1-n_{j,\mrm{p}}\right)+H$.
With perturbations $J_1 \sim J_2 \ll J_\perp$, we consider magnetic fields $|J_\perp-H|\sim J_1 \sim J_2\ll J_\perp \>(\sim H)$.
Since the number operator $n_{j,\mrm{p}}$ is zero or one, the chemical potential $\mu_{j,\mrm{m}}\gtrsim H$ is much greater than $\mu_{j,\mrm{p}}\sim 0$.
Thus, the energy scales of quasi-spins split off by a strong magnetic field, where the low-energy physics is described by the quasi-spin of `p'.
Since the magnetization process is obtained by the expectation value $\lag n_{j,\mrm{p}}-n_{j,\mrm{m}}\rag$ as a function of $H$ and $\lag n_{j,\mrm{m}}\rag\sim 0$, we can focus the discussion on the low-energy physics described by the quasi-spin of `p'.
To neglect the high-energy physics, we consider the effective Hamiltonian in a Hilbert space projected by $\mcl{H}_{\mrm{eff}}=\mcl{P}\mcl{H}\mcl{P}$ with $\mcl{P}=\prod_j (1-n_{j,\mrm{m}})$.
Here, the projected space has the expectation value of the number operator of the `m' quasi-spin $\lag n_{j,\mrm{m}}\rag = 0$.
In this approximation, the up and down states of the quasi-spin of `p' correspond a triplet and a singlet states on a bond, respectively (see also  Fig. 2). 
In the projected Hilbert space, we finally find a frustrated spin chain again, where spins are substituted with the quasi-spins as follows~\cite{giamarchi99,mila98,batista12},
\begin{equation}
\mcl{H}_{\mrm{eff}} = \mcl{P} (\mcl{H}_\parallel^\prime + \mcl{H}_Z^\prime )\mcl{P} \label{eq:ham1}
\end{equation}
with
\begin{eqnarray}
\mcl{H}_\parallel^\prime &=& \sum_{eta=1,2}\sum_j \Bigg[ {J_{\eta}^z}^\prime T_{j,\mrm{p}}^zT_{j+\eta,\mrm{p}}^z \\ \nonumber
&& \qquad \qquad  + \frac{ {J_{\eta}^x}^\prime}{2} (T_{j,\mrm{p}}^+T_{j+\eta,\mrm{p}}^-+T_{j,\mrm{p}}^-T_{j+\eta,\mrm{p}}^+ )\Bigg], \\
\mcl{H}_Z^\prime &=& - H^\prime \sum_j T_{j,\mrm{p}}^z,
\end{eqnarray}
where the XY- and Ising-components of quasi-exchange interactions are denoted by ${J_{\eta}^x}^\prime=J_{\eta}$ and ${J_{\eta}^z}^\prime=J_{\eta}/2$ with $\eta=1,2$, respectively. 
See the Appendix for more details, by which one can find possible extensions of the theory.
This model has several differences from the original spin model in the weak rung coupling limit (see Table \ref{tab:comp}).
One is anisotropy of the anti-ferromagnetic spin interactions, where every quasi-exchange interaction is XY-like, ${J_{\eta}^x}^\prime=2{J_{\eta}^z}^\prime$.
In addition, we should be careful on the meaning of the magnetization of up and down quasi-spins, because the transformation maps the singlet state to {\it down} and the triplet to {\it up} quasi-spins.
Thus, the total magnetization with $q$-filling of up spins is not $2(q-1/2)$ but $q$. 
It must be also noted that the quasi-spins couple to the quasi-magnetic field given by, $H^\prime = H - J_\perp - (J_1+J_2)/2$.
This causes the zero expectation value of number of up spins without the real magnetic field $H=0$, that is, the singlet states pack in rung bonds in the ground state.

\begin{center}
\begin{table*}
\caption{Comparison between two different limits.}
\label{tab:comp}
\begin{tabular}{l|cc}
    & Weak rung-coupling limit & Strong rung-coupling limit \\
\hline
  Effective model & Decoupled frustrated spin-$\frac{1}{2}$ chains & Frustrated quasi-spin-$\frac{1}{2}$ chain\\
  Spin interaction & Heisenberg & XY-like \\
  Spin operator & $\bm{S}_{j,\mrm{u}(\mrm{l})}$ & $\bm{T}_{j,\mrm{p}}$ \\
  $M/M_{\mrm{sat}}$ for $q$-filling of up spins & $2(q-1/2)$ & $q$ \\
  Effective magnetic field & $H$ & $H^\prime = H - J_\perp - (J_1+J_2)/2 $ \\
  Filling of up spins at $H=0$ & $1/2$ & $0$ 
\end{tabular}
\end{table*}
\end{center}

With the help of the quasi-spin model, we can approach the natures of the magnetization process in the strong rung-coupling limit.
Firstly, the magnetization $M/M_{\mrm{sat}} =1/2$ with the strong rung coupling corresponds to $M/M_{\mrm{sat}} =0$ in the weak rung-coupling limit.
The gapped ground state in the weak rung-coupling limit is understood by the Majumdar-Gosh state with spontaneously-broken translational symmetry, where the spin singlet dimers pack over the legs.
In the same manner, the quasi-spin singlet dimers pack over the legs, although the magnetic moment $M=+1$ is set on a singlet dimer in Fig.~\ref{fig:plateaux}(b).
The 1/3 magnetization plateau with the weak rung coupling is equivalent to $2/3(=1/3\times1/2+1/2)$ plateau in the strong rung-coupling limit.
The freezing-spin structure and long-ranged order are also found in the strong rung-coupling limit, where the real spins of the upper and lower chains coherently freeze.
Additionally, we can find a $1/3(=-1/3\times1/2+1/2)$ plateau and some cusps in the strong rung-coupling limit, which correspond to the $M/M_{\mrm{sat}}=-1/3$ plateau and cusp singularities with low magnetic fields in the weak rung-coupling limit, respectively.

Therefore, the plateaux in the strong rung coupling correspond to $M/M_{\mrm{sat}}=\pm 1/3$ and $M/M_{\mrm{sat}}=0$ plateaux of the frustrated spin chain, although physical roles are played by the quasi-spins (see Fig.~\ref{fig:plateaux}(b)).
Actually, we confirm that the spins freeze coherently between the upper and lower chains in the $1/3$- and $2/3$-plateau states, and that quasi-spin correlation rapidly decays in the $1/2$-plateau state. 
Moreover, we can see good correspondence between Fig.~\ref{fig:mh}(a) with the weak rung-coupling and the quarter area at the upper right in  Fig.~\ref{fig:mh}(c) with the strong rung-coupling including a cusp singularity. 
Since the filling of  the quasi-up-spins with the strong rung coupling corresponds to that in the ground state without a magnetic field in the weak rung coupling limit, the magnetization curve in Fig.~\ref{fig:mh}(a) is folded in half in Fig.~\ref{fig:mh}(c). 
Thus, we conclude that the reconstruction of the quasi-particles results in the emergence of the isostructural magnetization curves in two different limits.

\section{Summary}
\label{sec:sum}
In summary, we have studied the magnetization curve of the frustrated spin-ladder system using the DMRG method, and the three magnetization plateaux are found at $M/M_{\mrm{sat}}=1/3$, $1/2$, and $2/3$ due to frustration. 
In the limit of strong-rung coupling, the magnetization near the field-induced phase transition ($H\sim H_{c1}$) can be understood by the triplons on rungs as hard-core bosons. 
On the other hand, the plateaux in $H_{c1} < H < H_{c2}$ can be explained by the quasi-spinons, which are reconstructed from the singlet and the triplet of spin pairs on a rung. 
The plateau at $M/M_{\mrm{sat}}=1/2$ corresponds to the valence bond solid of the quasi-spinons, while the plateaux at $1/3$ and $2/3$ can be associated with the array of  quasi-spinons such as soliton lattice. 
This is different from the usual BEC picture of triplons. 
The magnetization curves around $1/3$ and $2/3$ look like to be folded in half with same characteristics. 
This is also naturally explained using the effective model of quasi-spinons with spontaneously-broken symmetry. 

Our results will be useful for frustrated quantum spin ladder systems such as BCPO.
Concerning BCPO, it is noted that an anisotropic interaction, i.e., Dzyaloshinsky-Moriya (DM) interaction, is reported by the inelastic neutron scattering study~\cite{plum14}. 
The DM interaction also change the usual BEC picture of triplons on the magnetization process in the spin ladder system~\cite{sirker04}. 
The transition will be modified such as the second order one due to the DM interactions, which lead to linear terms of triplons. 
When the DM interaction is smaller than the rung coupling, our picture of the quasi-spinons on the magnetization curve will not be changed. 
Effects of such an interaction on the magnetic plateaux will be clarified in the near future. 

\begin{acknowledgments}
We would like to thank M. Fujita and O. P. Sushkov for valuable discussions. 
This work was partly supported by Grant-in-Aid for Scientific Research (Grant No.25287094, No.26103006, No.26108716, No.26247063, No.26287079, No.15H03553, and No.15K05192), by bilateral program from MEXT, by MEXT HPCI Strategic Programs for Innovative Research (SPIRE) (hp140215) and Computational Materials Science Initiative (CMSI), and by the inter-university cooperative research program of IMR, Tohoku University.
Numerical computation in this work was carried out on the supercomputers at JAEA, the K computer at the RIKEN Advanced Institute for Computational Science, and the Supercomputer Center at Institute for Solid State Physics, University of Tokyo.
\end{acknowledgments}

\appendix
\section{Derivation of effective Hamiltonian and quasi-spin operators}
In this section, we present a derivation of effective Hamiltonian with a quasi-spin transformation in the strong rung-coupling limit, $J_1/J_\perp \ll 1$ and $J_2/J_\perp\ll 1$.
With a finite magnetization in this limit, a quasi-spin transformation and a reduced Hamiltonian are useful to understand the ground state and the low-energy physics.
In order to obtain the reduced Hamiltonian, we start with two spin Hamiltonian on $j$-th rung as follows,
\begin{eqnarray}
\mcl{H}_j^{\mrm{rung}}&=&J_\perp\bm{S}_{j,\mrm{u}}\cdot\bm{S}_{j,\mrm{l}}-H\sum_{i=\mrm{u,l}}\bm{S}_{j,i} \nonumber\\
&=&\frac{J_\perp}{2} (d_{j,\mrm{u}}^\dagger d_{j,\mrm{l}} + d_{j,\mrm{l}}^\dagger d_{j,\mrm{u}}) + J_\perp n_{j,\mrm{u}}n_{j,\mrm{l}}\nonumber\\
& & -\left(\frac{J_\perp}{2} +H \right)(n_{j,\mrm{u}} + n_{j,\mrm{l}})+\frac{J_\perp}{4}-H,
\end{eqnarray}
with a Jordan-Wigner transformation of spin operators,
\begin{equation}
d_{j,\mrm{u}} = S_{j,\mrm{u}}^- e^{- i\frac{\pi}{2}(S_{j,\mrm{l}}+\frac{1}{2})}, \hspc d_{j,\mrm{l}} = S_{j,\mrm{l}}^- e^{+ i\frac{\pi}{2}(S_{j,\mrm{u}}+\frac{1}{2})},
\end{equation}
and the number operators $n_{j,\mrm{u}(\mrm{l})}=d_{j,\mrm{u}(\mrm{l})}^\dagger d_{j,\mrm{u}(\mrm{l})} = S_{j,\mrm{u}(\mrm{l})}^z+\frac{1}{2}$.
As it is well-known, these Jordan-Wigner operators obey anti-commutation relation, $\{d_{j,i}, d_{j,k}^\dagger\}=\delta_{i,k}$ and $\{d_{j,i}, d_{j,k}\}=\{d_{j,i}^\dagger, d_{j,k}^\dagger\}=0$.
To diagonalize this Hamiltonian, we can use a bonding and an anti-bonding operators for create and annihilate operators of Jordan-Wigner fermions, $d_{j,\mrm{b}} = (d_{j,\mrm{u}} + d_{j,\mrm{l}})/\sqrt{2}$, $d_{j,\mrm{a}} = (d_{j,\mrm{u}} - d_{j,\mrm{l}})/\sqrt{2}$. 
With the number operator of the bonding and an anti-bonding Jordan-Wigner fermions $n_{j,\mrm{b}(\mrm{a})}$, the rung Hamiltonian is rewritten as,
\begin{equation}
\mcl{H}_j^{\mrm{rung}} = n_{j,\mrm{a}}(n_{j,\mrm{b}}-1)-h(n_{j,\mrm{a}}+n_{j,\mrm{b}}-1) +\frac{1}{4}.
\end{equation}
We can obtain a quasi-spin transformation Eqs.~(\ref{eq:top1}) and (\ref{eq:top2}) with an inverse Jordan-Wigner transformation as follows,
\begin{equation}
T_{j,\mrm{p}}^+ = id_{j,\mrm{b}}^\dagger, \hspc T_{j,\mrm{p}}^- = -i d_{j,\mrm{b}}, \hspc T_{j,\mrm{p}}^z= n_{j,\mrm{b}}-\frac{1}{2},
\end{equation}
and
\begin{equation}
T_{j,\mrm{m}}^- = d_{j,\mrm{a}}^\dagger e^{-i\pi n_{j,\mrm{b}}}, \hspc T_{j,\mrm{m}}^+ = d_{j,\mrm{a}} e^{i\pi n_{j,\mrm{b}}}, \hspc T_{j,\mrm{m}}^z = \frac{1}{2} - n_{j,\mrm{a}}.
\end{equation}
These operators also satisfy the spin SU(2) algebra for themselves, and commutate each other.
The quasi-spin operators rewrite the rung Hamiltonian as,
\begin{eqnarray}
\mcl{H}_j^{\mrm{rung}}&=& -J_\perp \left(T_{j,\mrm{p}}^z-\frac{1}{2}\right)\left(T_{j,\mrm{m}}^z-\frac{1}{2}\right)-H(T_{j,\mrm{p}}^z-T_{j,\mrm{m}}^z) +\frac{J_\perp}{4} \nonumber\\
&=& -H_{\mrm{p}} T_{j,\mrm{p}}^z +  H_{j,\mrm{m}} T_{j,\mrm{m}}^z,  
\end{eqnarray}
where effective quasi-magnetic fields,
\begin{equation}
H_{\mrm{p}} \equiv H -\frac{J_\perp}{2}, \hspc H_{j,\mrm{m}} \equiv H +\frac{J_\perp}{2} - T_{j,\mrm{p}}^z > H.
\end{equation}
With the quasi-spin operators, the leg Hamiltonian between $j$-th and $(j+L)$-th rungs is given by,
\begin{widetext}
\begin{eqnarray}
\mcl{H}_{j,\eta}^{\mrm{leg}} &=& J_\eta\sum_{i=\mrm{u,l}} \bm{S}_{j,i}\cdot\bm{S}_{j+L,i} \nonumber\\
&=& \frac{J_\eta}{2} \left(T_{j,\mrm{p}}^z-T_{j,\mrm{m}}^z\right)\left(T_{j+L,\mrm{p}}^z-T_{j+L,\mrm{m}}^z\right)-\left(T_{j,\mrm{p}}^+ T_{j,\mrm{m}}^+-T_{j,\mrm{p}}^- T_{j,\mrm{m}}^-\right)\left(T_{j+L,\mrm{p}}^+ T_{j+L,\mrm{m}}^+-T_{j+L,\mrm{p}}^- T_{j+L,\mrm{m}}^-\right) \nonumber \\
& & \htab +\frac{J_\eta}{2}\Bigg\{T_{j,\mrm{p}}^+ T_{j+L,\mrm{p}}^-\cos\left[ \frac{\pi}{2}(T_{j,\mrm{m}}^z-T_{j+L,\mrm{m}}^z)\right] +T_{j,\mrm{m}}^+ T_{j+L,\mrm{m}}^-\cos\left[ \frac{\pi}{2}(T_{j,\mrm{p}}^z-T_{j+L,\mrm{p}}^z)\right] \nonumber\\
& & \htab\htab - T_{j,\mrm{p}}^+ T_{j+L,\mrm{m}}^+ \cos\left[ \frac{\pi}{2}(T_{j,\mrm{m}}^z-T_{j+L,\mrm{p}}^z)\right] -T_{j,\mrm{m}}^- T_{j+L,\mrm{p}}^-\cos\left[ \frac{\pi}{2}(T_{j,\mrm{p}}^z-T_{j+L,\mrm{m}}^z)\right] +\mrm{H. c.}  \Bigg\}.
\end{eqnarray}
\end{widetext}
If we consider small quasi-magnetic field for $T_{j,\mrm{p}}$ spin, namely $|H_{\mrm{p}}/J_\perp| \ll 1$, the quasi-magnetic field for $T_{j,\mrm{m}}$ spin is much larger than $|H_{\mrm{p}}|$, namely $|H_{j,\mrm{m}}|\gtrsim \frac{J_\perp}{2}\gg |H_{\mrm{p}}|$.
To deal low-energy physics, we can project out the high-energy states, that is, quasi-up-spins of $T_{j,\mrm{m}}$ opeartors.
With the projection operator given by $\mcl{P}=\prod_j \left(T_{j,\mrm{m}}^z-\frac{1}{2}\right)$, an effective Hamiltonian is obtained as $\mcl{P}\mcl{H}\mcl{P}=\mcl{H}_{\mrm{eff}}$.
Since the original Hamiltonian (\ref{eq:ham0}) is composed by a sum over the rung and leg Hamiltonians, $\mcl{H}_j^{\mrm{rung}}$ and $\mcl{H}_{j,L}^{\mrm{leg}}$, the quasi-spin transformation gives us the effective Hamiltonian Eq.~(\ref{eq:ham1}).

\begin{figure}[ht]
\centering
\includegraphics[width=0.3\textwidth]{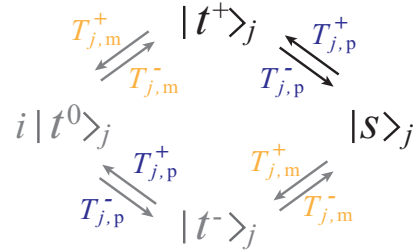}
\caption{(Color online) Schematic relationship between the quasi-spin operators and  the cofigurations of the singlet and triplets on $j$-th rung.
When the quasi-up-spin states of $T_{j,\mrm{m}}$ operators are projected out, the left-down side is prohibited. 
Thus, the quasi-up-spin and quasi-down-spin of $T_{j,\mrm{p}}$ operators approximately correspond to the triplet state $|t^+\rag_j$ and the singlet state $|s\rag_j$, respectively.}
\label{fig:top}
\end{figure}

We note that the quasi-spin operators can be written by singlet and triplets configurations on $j$-th rung as follows,
\begin{eqnarray}
T_{j,\mrm{p}}^z &=& \frac{1}{2} (|t^+\rag_j\lag t^+|_j + |t^0\rag_j\lag t^0|_j - |t^-\rag_j\lag t^-|_j -|s\rag_j\lag s|_j), \nonumber \\
T_{j,\mrm{p}}^+ &=& |t^+\rag_j\lag s|_j + i |t^0\rag_j\lag t^-|_j, \hspc \mrm{and}\hspc\mrm{H.c.},
\end{eqnarray}
and
\begin{eqnarray}
T_{j,\mrm{m}}^z &=& \frac{1}{2} (|t^-\rag_j\lag t^-|_j + |t^0\rag_j\lag t^0|_j -|t^+\rag_j\lag t^+|_j -|s\rag_j\lag s|_j),\nonumber \\
T_{j,\mrm{m}}^+ &=& |t^-\rag_j\lag s|_j + i |t^0\rag_j\lag t^+|_j, \hspc \mrm{and}\hspc\mrm{H.c.},
\end{eqnarray}
where $|t^\alpha\rag_j$ $(\alpha=\pm,0)$ and $|s\rag_j$ denote the triplets and singlet states on $j$-th rung, respectively.
When the quasi-up-spin states of $T_{j,\mrm{m}}$ operators are projected out, the quasi-up-spin and quasi-down-spin of $T_{j,\mrm{p}}$ operators approximately correspond to the triplet state $|t^+\rag_j$ and the singlet state $|s\rag_j$, respectively (see Fig.~\ref{fig:top}).

\begin{thebibliography}{00}
\bibitem{note1} The convergence of energy is confirmed up to a 72-rung system with the truncation number, $m_{\mrm{trunc}}\leq 500$.
\end{thebibliography}
\end{document}